\documentclass[12pt]{article}

\usepackage{cite}

\usepackage{amssymb,latexsym}

\usepackage{epstopdf}

\usepackage{graphics,epsfig}

\usepackage{color}

\let\a=\alpha \let\b=\beta \let\g=\gamma \let\d=\delta \let\e=\epsilon
  \let\th=\theta  \let\k=\kappa
\let\l=\lambda \let\m=\mu \let\n=\nu \let\x=\xi \let\p=\pi 
\let\s=\sigma   \let\f=\phi  
 
        \let\Th=\Theta \let\L=\Lambda
\let\X=\Xi  \let\S=\Sigma  \let\Y=\Psi
 
\let\la=\label  
  
\def\nn{\nonumber} \def\bd{\begin{document}} \def\ed{\end{document}}
\def\ds{\documentstyle} \let\fr=\frac \let\bl=\bigl \let\br=\bigr
\let\Br=\Bigr \let\Bl=\Bigl
\let\bm=\bibitem
\let\na=\nabla
\def\tU{{\widetilde U}}
\let\pa=\partial \let\ov=\overline
\def\ie{{\it i.e.\ }}
\newcommand{\be}{\begin{equation}}
\newcommand{\ee}{\end{equation}}
\def\ba{\begin{array}}
\def\ea{\end{array}}
\def\ft#1#2{{\textstyle{{\scriptstyle #1}\over {\scriptstyle #2}}}}
\def\fft#1#2{{#1 \over #2}}
\def\F#1#2{{ F_{#1}^{(#2)} }}
\def\cF#1#2{{ {\cal F}_{#1}^{(#2)} }}

\def\R{{\bf R}}
\def\sst#1{{\scriptscriptstyle #1}}
\def\oneone{\rlap 1\mkern4mu{\rm l}}
\def\e7{E_{7(+7)}}
\def\td{\tilde}
\def\wtd{\widetilde}
\def\im{{\rm i}}
\def\bog{Bogomol'nyi\ }
\newcommand{\ho}[1]{$\, ^{#1}$}
\newcommand{\hoch}[1]{$\, ^{#1}$}
\newcommand{\bea}{\begin{eqnarray}}
\newcommand{\eea}{\end{eqnarray}}
\newcommand{\ra}{\rightarrow}
\newcommand{\lra}{\longrightarrow}
\newcommand{\Lra}{\Leftrightarrow}
\newcommand{\ap}{\alpha^\prime}
\newcommand{\bp}{\tilde \beta^\prime}
\newcommand{\cB}{{\cal B}}
\newcommand{\cO}{{\cal O}}
\newcommand{\vecx}{\vec{x}}
\newcommand{\vecy}{\vec{y}}
\newcommand{\vecp}{\vec{p}}
\newcommand{\vecq}{\vec{q}}
\newcommand{\tr}{{\rm tr} }
\newcommand{\Tr}{{\rm Tr} }
\newcommand{\NP}{Nucl. Phys. }

\newcommand{\cL}{{\cal L}}
\newcommand{\cA}{{\cal A}}
\newcommand{\cT}{{\cal T}}
\newcommand{\cD}{{\cal D}}
\newcommand{\cH}{{\cal H}}

\def\sst#1{{\scriptscriptstyle #1}}
\def\0{{\sst{(0)}}}
\def\1{{\sst{(1)}}}
\def\2{{\sst{(2)}}}
\def\3{{\sst{(3)}}}
\def\4{{\sst{(4)}}}
\def\5{{\sst{(5)}}}
\def\6{{\sst{(6)}}}
\def\7{{\sst{(7)}}}
\def\8{{\sst{(8)}}}
\def\9{{\sst{(9)}}}
\def\p{{\sst{(p)}}}
\def\q{{\sst{(q)}}}
\def\ve{\varepsilon}
\def\vf{\varphi}
\def\F{\Phi}
\def\wg{\wedge}

\def\thb{\bar{\theta}}
\def\Thb{\bar{\Theta}}
\def\barp{\bar{p}}
\def\barq{\bar{q}}
\def\barc{\bar{c}}
\def\bard{\bar{d}}
\def\e{\epsilon}

\def \bi{\bibitem}
\def \la {\label}

\def \l {\lambda}
\def\foot{\footnote}
\def \tl  {{\tilde \l}}
\def \sql {{\sqrt \l}}
\def \adss {$AdS_5 \times S^5$\ }
\newcommand{\rf}[1]{(\ref{#1})}
\def \ov {\over}

\def\th{\theta}
\def\Th{\Theta}
\def\vth{\vartheta}
\def\btheta{{\bar\theta}}
\def\ttheta{{{\tilde\theta}}}
\def\bttheta{{{\bar\ttheta}}}
\def\vth{\vartheta}

\def\ra{\rightarrow}
\def\N{\nabla}
\def\F{{\cal F}}
\def\uM{\underline{M}}
\def\uA{\underline{A}}
\def\uN{\underline{N}}
\def\uP{\underline{P}}
\def\ua{\underline{a}}
\def\ub{\underline{b}}
\def\uc{\underline{c}}
\def\ud{\underline{d}}
\def\ue{\underline{e}}
\def\uf{\underline{f}}
\def\ui{\underline{i}}
\def\uj{\underline{j}}
\def\uk{\underline{k}}
\def\ul{\underline{l}}
\def\ual{\underline{\alpha}}
\def\ube{\underline{\beta}}
\def\um{\underline{m}}
\def\un{\underline{n}}
\def\up{\underline{p}}
\def\uq{\underline{q}}
\def\ur{\underline{r}}
\def\us{\underline{s}}
\def\umu{\underline{\mu}}
\def\unu{\underline{\nu}}
\def\ula{\underline{\l}}
\def\uka{\underline{\k}}
\def\usi{\underline{\s}}
\def\urh{\underline{\r}}
\def\cc{\circ}
\def\eqv{\equiv}

\def\ni{\noindent}

\def\Ep{E^{{}^{(+)}}}
\def\Em{E^{{}^{(-)}}}

\def\Mp{M^{{}^{(+)}}}
\def\Mm{M^{{}^{(-)}}}

\def \ha{{1\ov 2}}

\def\r{\rho}

\def\Y{{\rm Y}}
\def\X{{\rm X}}
\def\tY{\tilde{\rm Y}}
\def\tX{\tilde{\rm X}}
\def\dY{\dot{\rm Y}}
\def\dX{\dot{\rm X}}

\def \J {\mathcal{J}}
\def \del {\partial}

\def\dF{\dot{F}}
\def\dG{\dot{G}}
\def\df{\dot{f}}
\def \E {{\cal E}}
\def \S {{\cal S}}
\def \J {{\cal J}}

\def\ms{\mathcal{S}}
\def\mj{\mathcal{J}}
\def\soj{\fr{\ms}{\mj}}
\def \R {{\bf R}}
\def \om {\omega}
\def \bE {\bar E}
\def \x {{\cal X}}

\def \bi{\bibitem}
\def \la {\label}

\def \l {\lambda}
\def\foot{\footnote}
\def \tl  {{\tilde \l}}
\def \sql {{\sqrt \l}}
\def \adss {$AdS_5 \times S^5$\ }
\def \ov {\over}

\def \varpi {{\rm w}}

\def\thb{\bar{\theta}}
\def\Thb{\bar{\Theta}}
\def\mb{\bar{\m}}
\def\ab{\bar{\a}}
\def\zb{\bar{z}}
\def\psib{\bar{\psi}}
\def\barp{\bar{p}}
\def\barq{\bar{q}}
\def\barc{\bar{c}}
\def\bard{\bar{d}}
\def\e{\epsilon}
\def\wb{\bar{w}}
\def\lb{\bar{\l}}
\def\Jb{\bar{J}}
\def\Nb{\bar{N}}
\def\Zb{\bar{Z}}
\def\pab{\bar{\pa}}

\def\At{\tilde{A}}
\def\Bt{\tilde{B}}
\def\Ct{\tilde{C}}
\def\Dt{\tilde{D}}
\def\Et{\tilde{E}}
\def\Ft{\tilde{F}}
\def\Gt{\tilde{G}}
\def\Ht{\tilde{H}}
\def\Kt{\tilde{K}}
\def\Mt{\tilde{M}}
\def\Nt{\tilde{N}}
\def\Rt{\tilde{R}}
\def\at{\tilde{a}}
\def\bt{\tilde{b}}
\def\ct{\tilde{c}}
\def\dt{\tilde{d}}
\def\et{\tilde{e}}
\def\ft{\tilde{f}}
\def\htil{\tilde{h}}
\def\gt{\tilde{g}}
\def\nt{\tilde{n}}
\def\mut{\tilde{\mu}}
\def\nut{\tilde{\nu}}
\def\pht{\tilde{\f}}
\def\vft{\tilde{\vf}}

\def\rht{\tilde{\rho}}

\def\asth{\hat{*}}
\def\phh{\hat{\phi}}

\def\bA{{\bf A}}

\def\ola{\overleftarrow}
\def\ora{\overrightarrow}
\def\alt{\tilde{\a}}

\def\eh{\hat{e}}
\def\Eh{\hat{E}}
\def\gh{\hat{g}}
\def\eph{\hat{\e}}
\def\ph{\hat{p}}
\def\alh{\hat{\a}}
\def\beh{\hat{\b}}
\def\gah{\hat{\g}}
\def\Fh{\hat{F}}
\def\muh{\hat{\m}}
\def\nuh{\hat{\n}}
\def\thh{\hat{\th}}
\def\rhh{\hat{\r}}
\def\dh{\hat{d}}
\def\ih{\hat{i}}
\def\jh{\hat{j}}
\def\hh{\hat{h}}
\def\kh{\hat{k}}
\def\rh{\hat{r}}
\def\tht{\hat{t}}
\def\deh{\hat{\d}}
\def\wh{\hat{w}}
\def\lah{\hat{\l}}
\def\Ah{\hat{A}}
\def\Kh{\hat{K}}
\def\Rh{\hat{R}}
\def\Ch{\hat{C}}
\def\Omh{\hat{\Omega}}

\def\xh{\hat{x}}

\def\ps{\rlap{\, /}\;\,p }
\def\ks{\rlap{\, /}\;\,k }

\def\gym{g_{YM}}

\def\adot{\dot{a}}
\def\bdot{\dot{b}}
\def\bpa{\bar{\pa}}

\def\pr{\prime}
\def\ssk{\medskip}
\def\clb{\color{blue}}
\def\clr{\color{red}}
\def\clg{\color{green}}

\begin{document}

\overfullrule=0pt
\parskip=2pt
\parindent=12pt
\headheight=0in \headsep=0in \topmargin=0in
\oddsidemargin=0in

\vspace{ -3cm}
\thispagestyle{empty}

 \vspace{0.1cm}

\setcounter{equation}{0}
\setcounter{footnote}{0}
\setcounter{section}{0}

\begin{center}

{\Large\bf Indication for unsmooth horizon induced by quantum gravity interaction}

\vskip 0.8cm

 \vspace{.5cm}

\vspace{0.5cm}
I. Y. Park
\\
\vspace{0.3cm}
{\it Department of Physics, Hanyang University\\
           Seoul 133-791, South Korea }

\vspace{0.3cm}

{\it Center for Quantum Spacetime, Sogang University\\
Shinsu-dong 1, Mapo-gu, 121-742 South Korea \\
}

\vspace{0.3cm}
{\it Department of Applied Mathematics,
Philander Smith College 
                               \\
Little Rock, AR 72223, USA \\
inyongpark05@gmail.com
}

\end{center}

 \vspace{0.1cm}

 \begin{abstract}

\ni The angular ADM reduction of the BTZ spacetime yields a Liouville-type theory.
The analysis of the resulting Liouville theory naturally
 leads to identification of the stretched horizon. The dynamics associated with the stretched horizon has a feature that seems consistent with the unsmooth horizon; the quantum gravity effects are essential for the unsmoothness. We show that the ``anomaly" term in the stress-energy tensor is responsible for the Planck scale energy experienced by an infalling observer.

\end{abstract}
\newpage

\section{Introduction}

Given the unrenormalizability of the 4D Einstein-Hilbert action, the semi-classical description has been widely used in the black hole literature. (See, e.g., \cite{Birrell} and \cite{muk} for reviews.)
This description led to the discovery of Hawking radiation and many other useful results, and we believe it is essential to go beyond it to solve Black Hole Information (BHI) problem and surrounding issues.

The validity of the semi-classical description is one of the four postulates of black hole complementarity (BHC) \cite{Susskind:1993if}. However,
 the postulates' mutual compatibility has been questioned in the recent work of \cite{Almheiri:2012rt}, according to which one or more of the four postulates of BHC must break down. In particular, an infalling observer will experience a firewall when crossing the event horizon of a sufficiently old black hole in `violation'\footnote{What is being challenged might not be Equivalence Principle itself but the conventional lore that Equivalence Principle implies a smooth horizon. We will come {back} to this in the conclusion.} of the Equivalence Principle.
 In this work, we analyze a 2D theory to study the behavior of the stretched horizon as observed by an infalling observer.

The 2D theory that we consider describes the quantum fluctuations of the selected 2D hypersurface of the BTZ spacetime \cite{Banados:1992wn}.\footnote{Recent works that tackle black hole information problem by analyzing concrete models include \cite{Kawai:2013mda,Berenstein:2013tya,Silverstein:2014yza}.}  (See \cite{Carlip:1998qw} for review of the BTZ black hole.) It was proposed in \cite{Hatefi:2012bp} that it should be possible to tackle the BHI problem without directly dealing with the quantization issue of the 3D or 4D gravity. One may tackle the BHI after first reducing the theory to a selected 2D hypersurface through the procedure called dimensional reduction to a hypersurface of foliation (or ADM reduction for short).

The ADM reduction is a variation of (but significantly different from) the standard Kaluza-Klein reduction.
The procedure was motivated by the endeavor to derive AdS/CFT from the first principle. To be specific, let us take the prime example, AdS$_5$/CFT$_4$.
The CFT, ${\cal N}=4$ SYM in this case, can be viewed as the theory of the hypersurface of AdS$_5$ at $r=\infty$.
By applying the procedure in the Hamilton-Jacobi formulation, it was shown that the 5D
AdS gravity admits a class of solutions with a ``moduli field", which in turn was identified as the abelian worldvolume (i.e., the hypersurface at a fixed $r$) gauge field \cite{Sato:2002kv}\cite{Hatefi:2012bp}. This may be viewed as the way in which the actual dualization of the bulk theory to the boundary theory should generally work.\footnote{ With the result in \cite{GonzalezRey:1998uh} and conjecture in \cite{Schwarz:2013wra} combined, the derivation of AdS$_5$/CFT$_4$
seems within close reach. The author thanks H.-S. Yang for discussion on this point.}

 The ADM reduction has subsequently been applied to 4D Einstein-Hilbert action \cite{Park:2013iqa}\cite{Park:2013vpa} and 3D AdS gravity action \cite{Park:2013bma}. In this work, we build on the latter case. There has been expectation in the literature \cite{Coussaert:1995zp,Frolov:1999my,Krasnov:2000ia,Giacomini:2003cg,
Chen:2003si,Yuan:2011gq,Nakatsu:1999wt} that the BTZ geometry should be associated with Liouville theory \cite{Belavin:1984vu}.\footnote{{As well known, 3D pure gravity does not have propagating degrees of freedom. Then there should be two possibilities regarding what is responsible for the Liouville degrees of freedom. Firstly, it could be non-perturbative degrees of freedom. Secondly, they could be associated with some type of "boundary" degrees of freedom. Out work is in line with the second possibility as we further contemplate in the conclusion.} }
 This expectation has been confirmed in the ADM setup \cite{Park:2013bma}: the theory reduced along the $\vf$ (respectively, $r$) direction has turned out to be curved-space Liouville (respectively, `flat-space' Liouville) theory.  (Liouville theory was obtained in \cite{Solodukhin:1998tc} by the standard dimensional reduction in the 4D Einstein-Hilbert case.)

In this work, we take the Liouville theory resulting from the $\vf$-reduction and analyze its implications for black hole physics. In particular, we study its implications for the behavior of the surface that we identify as the stretched horizon. We compute the zero-mode contribution to the expectation value of the 2D stress-energy tensor. Although the Liouville theory is super-renormalizable, various issues pertaining to detailed renormalization procedure are still present. We outline the procedure for computing the rest of the contributions (i.e., nonzero-mode contributions).

In order to examine the behavior of the stretched horizon, we consider the coordinate-invariant quantity that involves the 2D stress-energy tensor $T^S_{ab},\; a=(t,r)$:
\bea
<S|T^S_{ab}|S> U_S^a U_S^b  \la{mo}
\eea
where `$S$' denotes the Schwarzschild coordinates and $U_S^a$ is the timelike geodesic.
(A recent firewall-related analysis of the vacuum expectation value of the stress tensor appeared in \cite{Lowe:2013zxa}. See, e.g., \cite{Unruh:1976db,Fulling:1977zs,GP}\cite{Birrell} for early discussions.)
This quantity is the energy density as measured by an infalling observer.
As wellknown, Liouville theory has a central charge.
We show that the central charge term induces the invariant quantity above to generate the Planck scale energy density at the stretched horizon, a behavior consistent with the idea of the unsmooth horizon \cite{Braunstein:2009my}\cite{Almheiri:2012rt}.

\vspace{.3in}

The rest of the paper is organized as follows. In section 2, we review the $\vf$-reduction of the BTZ spacetime.
In section 3, we compute the zero-mode contribution to \rf{mo}.
Conceptually, narrowing down to the zero-mode contribution means an additional ADM reduction along the $r$ direction, and one gets to deal with the quantum mechanics of the `point stretched horizon', i.e., stretched horizon of zero spatial dimension.
Once the zero-mode result gets contracted with the {timelike} geodesic, one gets the Planck scale energy. The computation of the genuine 2D contributions is outlined. We end with discussion and future directions.

\section{Review of $\vf$-reduction}

In the next section which is the main body of the work, we identify the stretched horizon and analyze its dynamics by computing the zero-mode contribution to the
vacuum expectation value of the 2D stress-energy tensor.
To set the stage for the next section, we obtain the 2D theory by applying the ADM reduction technique along the angular direction \cite{Park:2013bma}.\footnote{Here we do not carefully trace the terms that arise from the virtual boundary contribution because they will not play a role in the 2D stress-energy tensor. (Such terms were carefully dealt with in \cite{Park:2013vpa} and \cite{Park:2013bma}.)}

\vspace{.2in}


Let us consider the 3D action
\be
S=\int\, d^3x\, \sqrt{-\gt^\3}\left[\Rt^\3+\fr2{l^2} \right]
\la{AdS3act}
\ee
where $l$ is the AdS length scale.
By employing the ADM formalism, the metric can be put in the form,
\bea
ds_3^2=(\nt_\vf^2+\htil^{ab}\Nt_a \Nt_b)d\vf^2+2\Nt_a d\vf dx^a+\htil_{ab}dx^adx^b,
\quad a=t,r
\eea
In this formalism, the 3D action takes
\bea
S&=&\int  d^2x  d\vf \, \sqrt{-\htil}\;\nt_\vf\left[\Rt^\2+\Kt^2-\Kt_{ab}\Kt^{ab}
 -2\L\right] \la{3drescorphi}
\eea
$\Nt_a$ can be {gauge-fixed} to $\Nt_a=0$.
Let us rescale the 3D metric $\htil_{\m\n}$ (where $\m=(a,\vf)=(t,r,\vf)$) by
\bea
\htil_{\m\n}=e^{2\rht(t,r,\vf)}h_{\m\n}
\eea
The field $\rht$ is related to the lapse function associated with $h_{\m\n}$: $n_\vf$ is defined as ${{n}}_\vf(t,r,{ \vf}) \equiv e^{{ -}\r(t,r)}$ {where $\r(t,r)$ is a dimensionally reduced $\tilde{\r}(t,r,\vf)$ field.}
This leads to
\bea
S=\int  d^2x  d\vf \, \sqrt{-h_\2}\;\Big[R^\2+K^2-K_{ab}K^{ab}
 -4{\N}_\3^2 \rht-2({\N}_\3 \rht)^2+\fr2{l^2}e^{2\rht}
  \Big] \la{3drescorphi3}
\eea
After reduction to 2D and by
setting $\rht(t,{ r},{ \vf}) =\r_0(r)+  \r(t,r)$ with {$e^{\r_0(r)}=r$}, one gets
\bea
S=\int  dtdr    \sqrt{-h_\2}\;{r}\left[R^\2
 +\a_\vf(r) \r  -2(\N_a\r)^2+(K^2-K_{ab}K^{ab})
 +\fr{2{r^2}}{l^2} e^{2\r}\right] \la{3drescor3q}
\eea
where $\a_\vf(r)=4\N^2\r_0$. $\a_\vf(r)$ can be adjusted to our needs by renormalization procedure. (See below \rf{3drescor3q3}.)
Let us gauge-fix the 2D metric $h_{\2 ab}$ to
\bea
ds^2_2=\tilde{\g}_{0ab} dx^a dx^b
\la{adsansr2q2}
\eea
where
\bea
\tilde{\g}_{0ab}\equiv
\left(
\begin{array}{cc}
 -{\fr1{r^2}}f(r) & 0  \\
 0 & \fr{1}{{ r^2}f(r)} \\
\end{array} \la{gamz2q}
\right)
\eea
Let us rescale the 2D metric of \rf{3drescor3q} to the original 2D part of the 3D metric so that
\bea
\tilde{\g}_{0ab}\equiv \fr{1}{r^2}\g_{ab},\quad\g_{0ab}=
\left(
\begin{array}{cc}
 -f(r) & 0  \\
 0 & \fr{1}{f(r)} \\
\end{array} \la{gamz2q2}
\right)
\eea
The action now takes
\bea
S&=&\int  d^2x    \sqrt{-\g_0}\left[
   -2{ r}(\N_a\r)^2+\fr{\a_\vf(r)}{ r} \r
 +\fr{2{ r}}{l^2} e^{2\r}\right] \la{3drescor3q2}
\eea
where the field-independent terms have been removed.
This can be rewritten
\bea
S&=&\int  d^2x    \left[
   \fr{{ r}}{8\pi f}(\pa_t\r )^2-\fr{{ r}f}{8\pi}(\pa_r\r )^2
+\fr{\a_\vf(r)}{16\pi { r}} \r +\fr{{ r}}{8\pi l^2} e^{2\r}\right] \la{3drescor3q3}
\eea
where the action has been numerically rescaled. The linear term may be omitted as part of renormalization procedure (namely, by the freedom to choose the starting action in renormalization procedure) \cite{Distler:1988jt}\cite{Seiberg:1990eb}\cite{Park:2013iqa}.
(Alternatively, the linear term does not appear at all if one rescales the 3D metric twice separately, first with $e^{2\r}$ and next with $e^{2\r_0}$.)

\section{Behavior of horizon }

To examine the behavior of the horizon (or, more precisely speaking, the stretched horizon to be identified below), let us consider the following coordinate invariant quantity: the energy density measured by a free-falling observer, given by
\bea
 <S|\;T_{ab}^S \;|S>U_S^a U_S^b \la{2dse}
\eea
Recall that the stress-energy tensor has a constant term that originates from the presence of the exponential interaction. The constant term plays an important role: the $<S|\;T_{ab}^S \;|S>$ term would vanish without it.


The geodesic, $U_S^a$, was worked out in \cite{Cruz:1994ir}; the timelike geodesic is given by
\bea
r^2\dot{r}^2 &=& -(r^4-r^2)+c_0^2 r^2  \la{rg}
\eea
\bea
\dot{t} &=& \fr{c_0 r^2}{r^2(r^2-r_H^2)} \la{tg}
\eea
where $c_0$ is an integration constant.

\subsection{1D description of stretched horizon}

Before we get to outline the 2D perturbative computation of \rf{2dse}, let us narrow down to the ground state sector of the theory. This sector pertains to the zero modes, $(q,p)$, of the mode expansion of {$\r$} and its canonical momentum $\Pi$. The mode expansion should take a form analogous to the flat case \cite{Seiberg:1990eb}\cite{Ginsparg:1993is}:
\bea
\r(t,r)&=&q(t)+\sum_{n\neq 0}\fr{i}{n}(a_n(t)e^{-inr^*}+b_n(t)e^{inr^*})\nn\\
\Pi(t,r)&=&p(t)+\sum_{n\neq 0}\fr{1}{4\pi}(a_n(t)e^{-inr^*}+b_n(t)e^{inr^*})
\eea
where $r^*$ is a tortoise-type coordinate \cite{Banados:1992gq}.
The zero-mode system may also be viewed as the 1D system that results from the additional radial reduction of \rf{3drescor3q3}:
\bea
S&=&\int  dt   \left[
   \fr1{8\pi }(\pa_tq )^2
 +\fr{{f}}{8\pi l^2} e^{2q}\right] \la{1Dact}
\eea
where we have rescaled the action by $f/r$. By choosing the location of the hypersurface at $r=\r_H+\d_P$ where $\d_P$ is a Planck scale distance, the $r$-reduction naturally leads to the notion of the stretched horizon degrees of freedom. In other words, the resulting 1D field $q(t)$ should be interpreted as the degree of freedom of the stretched horizon.

The full non-perturbative contribution from the ground sector can be found by solving the 1D Schrodinger equation with the Hamiltonian given by
\bea
H_0=\fr12\Big(\fr{\pa}{\pa q}\Big)^2+\fr{{ f} }{8l^2}e^{2 q}+\fr{1}{8} \la{zh}
\eea
It can be noted that the sign of the kinetic term has flipped when it is compared with the corresponding sign in \cite{Seiberg:1990eb}. This change originates from the ``wrong" sign of the kinetic term in \rf{3drescor3q3}.
The Schrodinger equation $H_0\psi=E\psi$ takes
\bea
\left[\Big(Z\fr{\pa}{\pa Z}\Big)^2 +\fr{{ f}}{4l^2}Z^2+\fr{1}{4} \right]\psi_e=
\Big(\fr{1}{4}+e\Big)\psi_e  \la{scheq}
\eea
where $Z\equiv e^{ q}$. The energy $E$ is $E=\fr{1}{8}+\fr{e}{2}$ and the solution is a Bessel function. (In the case of \cite{Seiberg:1990eb} the solution was a modified Bessel function.)
{For the lowest energy state in the ground sector, let us set $e=0$.}
 The Hamiltonian of \cite{Seiberg:1990eb} was positive-definite whereas the Hamiltonian \rf{zh} is not. Therefore, the energy, in particular, the value of $e$ can be negative in the present case. (See \cite{Kim:2013caa} for the recent discussion on appearance of negative energy in the context of Firewall.) We interpret
the appearance of the negative-$e$ states as a signal of instability \cite{Park:2013iqa}, and
will put the negative energy states aside. (An indication that this is justified is that in the the positive definite case of \cite{Seiberg:1990eb}, the negative energy case $e<0$ yields solutions that do not satisfy the boundary
condition. We are simply viewing the negative-$e$ states as associated with
the instability although the solutions satisfy the boundary condition in the present case.) Relatedly, with the positive-definite Hamiltonian, the energies of the $r$-dependent states will be higher than those of the $r$-independent states, and this pattern should remain true even in the present case once the negative energy sector is put aside.

For the interacting theory, the contribution of the zero-modes to the energy density as measured by the infalling observer should be
\bea
H_0 UU
\eea
where $U=\dot{t},\; H_0= T$ (the 1D stress-energy tensor) and $\dot{t}$ is given in \rf{tg}. For $r=r_H+\d_P$ it becomes
\bea
\dot{t} \simeq \fr{c_0 r_H^2}{2r_H^3}\fr1{\d_P}
\eea
 The Planck scale behavior\footnote{Observations on diverging stress-energy at the event horizon were made in the past as well \cite{Birrell}. The present work may therefore be viewed to some extent as reinterpretation of those observations.} comes from multiplication with $U$:
 \bea
 H_0 UU 
    \simeq \fr{c_0^2}{32r_H^2}\fr1{\d_P^2}
 \eea
If there were no interaction term, this contribution could not arise:
for the branch of positive energy, $E>0$, the solution in the absence of the $Z^2$ term in \rf{scheq} diverges as $l\ra \infty$ and should be abandoned. We take this as an indication that the
free theory does not display the Planck scale energy at the stretched horizon.

In passing, let us also note the stress-energy measured by the
stationary Schwarzschild observer,
\bea
 <S|\;T_{00}^S \;|S>
\eea
 The zero-mode contribution should be just 1/8 according to the computation
above. (The expected $r$-dependence will appear once the nonzero mode contributions are taken into account.)


\subsection{2D approach}

It should be possible to carry out the perturbative analysis by treating the exponential term as a perturbation.
The loop divergences are expected and renormalization should be carried out.
As part of the renormalization procedure, the starting point of the 2D action can be taken as \rf{3drescor3q3}, which we quote here for convenience (the linear term has been omitted):
\bea
S&=&\int  d^2x    \left[
   \fr{{ r}}{8\pi f}(\pa_t\r )^2-\fr{{ r}f}{8\pi}(\pa_r\r )^2
 +\fr{1}{8\pi l^2} e^{2\r}\right] \la{3drescor3q3q}
\eea
Once the renormalization is complete, one can compute the vacuum
expectation value of
\bea
T_{\pm \pm}&=& -\fr12 (\pa_\pm \r)^2
        +\fr{{ 1}}{8l^2}e^{2 \r}+\fr{1}{8}
\eea
in some appropriately redefined 2D coordinates.
This computation will involve construction of the Green function among other things, and to that end it will be useful to make use of the Kruskal-type coordinates obtained in \cite{Banados:1992gq}\cite{Carlip:1995qv} in the intermediate step. The advantage for employing the Kruskal type coordinates is clear: the Laplace equation takes the same form as the flat case.
It should be possible to express the Green function in terms of the sum over the mode functions. Then the stress-energy tensor can be computed along the lines of the corresponding flat space computation.

\section{Conclusion}

In this work, we have computed the ground state contribution to the energy density as measured by a free-falling observer in the 2D theory. The additional ADM reduction
along $r$ has led to the natural identification of the stretched horizon.
The ground sector is spanned by the zero modes of the field $\r$ and its conjugate momentum $\Pi$. We have shown that the quantum gravity interaction is essential for the stretched horizon to display the Planck scale excitations observed by a free-falling observer.
The result of this work supports the unsmooth horizon proposals in the literature. 
We believe that the result is at odds with the conventional lore -which is based on semi-classical physics -that Equivalence Principle implies smooth horizon (rather than Equivalence Principle itself).  

{In one of the footnotes, we have briefly addressed the origin of the Liouville degrees of freedom. The idea of the virtual boundary associated with ADM reduction seems to be in line with the proposal in \cite{Carlip:1999db} to view a horizon as a boundary. There is an interesting investigation that needs to be done to fully justify the view of a horizon as a boundary surface. Carrying out the ADM procedure, one will obtain a Liouville type theory even if one chooses the radial location to be outside of the horizon, i.e., a bulk point. As well known, 3D gravity does not have any propagating degrees of freedom. The Liouville theory in this case should not be taken 
as genuine degrees of freedom since it should be possible to remove all of its dof by gauge symmetry. However, if one chooses the radial location at the event horizon, it is expected that the gauging procedure would presumably encounter some type of singularities due to quantum effects. 
}

One obvious future direction is to compute the $r$-dependent sector's contribution to the stress-energy; renormalization procedure must be completed prior to this task.
Once one reaches this point it would be possible to see the additional role of the interaction: information would be coded on the stretched horizon in the manner that reflects the interaction.

Another direction is with regard to the teleological nature of an event horizon \cite{tele}. The teleological nature of the stretched horizon was discussed in \cite{Susskind:1993if}. We believe that the teleological nature
of the stretched horizon may well play a role in the proposed mechanisms of blackening and bleaching \cite{Park:2013rm}.

It should be possible to use the 2D setup that was reviewed in section 2 to study
the potential presence of a bleaching mechanism. For that, it would be useful to construct a wave-packet and follow its time evolution. Some of the ingredients in \cite{Schoutens:1993hu} will be useful. An interacting QFT approach that shares a certain spirit with the semi-classical description would be required; most textbooks on QFT focuses on computing S-matrix. In our view, the best setup should be what is called the Schrodinger approach of QFT \cite{Hatfield}.

\vspace{.2in}
We will report on some of these issues in the near future.

\vspace{.7 in}
\ni {\bf Acknowledgments}
\\

\ni I would like thank the members of CQUeST Sogang University, the physics department of Hanyang University, Kyung Hee University, and KIAS for useful discussions.
 I especially thank B.-H. Lee for his hospitality during my visit to CQUeST.

\newpage
\appendix

\renewcommand{\theequation}{A.\arabic{equation}}
\setcounter{equation}{0}

\newpage

\end{document}